# Designing wake-up free ferroelectric capacitors based on the $HfO_2/ZrO_2$ superlattice structure

*Na Bai,*[1] *Kan-Hao Xue,*[1,2*] *Jinhai Huang,*[1] *Jun-Hui Yuan,*[1] *Wenlin Wang,*[1] *Ge-Qi Mao,*[1] *Lanqing Zou,*[1] *Shengxin Yang,*[1] *Hong Lu,*[1] *Huajun Sun,*[1,2*] *and Xiangshui Miao*[1,2]

[1]School of Integrated Circuits, School of Optical and Electronic Information, Huazhong University of Science and Technology, Wuhan 430074, China

[2]Hubei Yangtze Memory Laboratory, Wuhan 430074, China

*Corresponding Authors. Email: xkh@hust.edu.cn (K.-H. Xue), shj@hust.edu.cn (H. Sun)

**Abstract**

The wake-up phenomenon widely exists in hafnia-based ferroelectric capacitors, which causes device parameter variation over time. Crystallization at higher temperatures have been reported to be effective in eliminating wake-up, but high temperature may yield the monoclinic phase or generate high concentration oxygen vacancies. In this work, a unidirectional annealing method is proposed for the crystallization of $Hf_{0.5}Zr_{0.5}O_2$ (HZO) superlattice ferroelectrics, which involves heating from the $Pt/ZrO_2$ interface side. Nanoscale $ZrO_2$ is selected to resist the formation of monoclinic phase, and the chemically inert Pt electrode can avoid the continuous generation of oxygen vacancies during annealing. It is demonstrated that 600°C annealing only leads to a moderate content of monoclinic phase in HZO, and the TiN/HZO/Pt capacitor exhibits wake-up free nature and a $2P_r$ value of 27.4 μC/cm$^2$. On the other hand, heating from the $TiN/HfO_2$ side, or using 500°C annealing temperature, both yield ferroelectric devices that require a wake-up process. The special configuration of $Pt/ZrO_2$ is verified by comparative studies with several other superlattice structures and HZO solid-state solutions. It is discovered that heating from the $Pt/HfO_2$ side at 600°C leads to high leakage current and a memristor behavior. The mechanisms of ferroelectric phase stabilization and memristor formation have been discussed. The unidirectional heating method can also be useful for other hafnia-based ferroelectric devices.

## I. Introduction

The ferroelectricity observed in hafnia ($HfO_2$)-based oxides, especially $Hf_{0.5}Zr_{0.5}O_2$ (HZO), has great revived the research interests for ferroelectric random access memory (FeRAM), ferroelectric field effect transistor (FeFET) as well as negative capacitance field effect transistor at the nanoscale dimension.[1-12] In contrast to traditional ferroelectrics such as perovskite $Pb(Zr,Ti)O_3$, $SrBi_2Ta_2O_9$ and organic polyvinylidene fluoride polymers, a special wake-up phenomenon has been widely observed in hafnia-based ferroelectric capacitors, which implies that the switchable polarization is subject to variation during the initial period of capacitor lifetime.[13] Although the spontaneous polarization is actually being enhanced during wake-up,[14] the variable device parameter can cause severe problems in applications. It has been reported that the crystallization temperature is highly relevant to the wake-up process.[13,15-17] In particular, a high crystallization temperature could effectively render wake-up-free ferroelectric capacitors based on HZO.[17] Therefore, raising the processing temperature is a potential solution to get rid of the wake-up phenomenon.

Nevertheless, using high crystallization temperature leads to other serious problems. On the one hand, hafnia-ferroelectric capacitors usually involve TiN or TaN electrodes, which are oxygen scavenger materials at high temperature.[18] On the other hand, high temperature crystallization tends to generate the thermodynamically stable monoclinic ($P2_1/c$) phase (*m*-phase) for $HfO_2$ or HZO, which are paraelectric.[16,19] The robust stability of monoclinic phase hinders its transformation into the ferroelectric orthorhombic $Pca2_1$ phase (*o*-phase). Hence, in most published works, the crystallization temperature was controlled at around 500°C. [17, 20, 21]

In this work, we attempt a special design for the ferroelectric layer that can avoid the above two problems, thus permitting a higher crystallization temperature. First of all, the concept of $HfO_2/ZrO_2$ superlattice enables crystallization from the $ZrO_2$ side. This is possible when $ZrO_2$ is adjacent to an inert electrode and the heat source is only from that electrode side during rapid thermal annealing (RTA). Standard RTA instruments typically possess heating modes from upper lamps, lower lamps and both. As is well-known, the monoclinic phase of $ZrO_2$ can be less stable than its tetragonal phase due to surface energy consideration, as long as the grain dimension is sufficiently small.[22-24] Hence, even high temperature RTA may not transform nanoscale $ZrO_2$ into *m*-phase. Secondly, an inert electrode like Pt can be selected in conjunction with the $ZrO_2$ layer that is closest to the heat source, in order to suppress the generation of new oxygen vacancies during crystallization annealing. In our experiments, the as-designed samples have been shown to possess the desired properties. Experimental details and mechanism analysis are given in the following sections.

## II. Wake-up Free Ferroelectric Capacitors through Design

Our prototype device structure is illustrated in Figure 1(a), with a TiN/$HfO_2$-$ZrO_2$ superlattice/Pt capacitor structure. The 12-nm $HfO_2$/$ZrO_2$ superlattice was grown on TiN through thermal atomic layer deposition (ALD). The 100 nm-thick Pt top electrode (TE) was deposited using DC sputtering. Detailed experimental parameters are specified in the *Experimental* section. The two interfaces formed by the ferroelectric layer with the TE and the bottom electrode (BE) therefore experienced quite different processes. The bottom interface went through a mild ALD process, but the top interface has experienced a high energy bombardment during sputtering. Such asymmetric treatment is quite common and recently it was reported that the abundant oxygen vacancies around the top interface may be related to the imprint phenomenon in HZO-ferroelectrics.[25] In our device, the impact of bombardment is only prominent in the topmost $ZrO_2$ layer of ~2 nm thick. The underneath $HfO_2$ layer is well maintained after TE sputtering, as evidenced from the scanning transmission electron microscopy (STEM) analysis in Figures 1(d)-1(e).

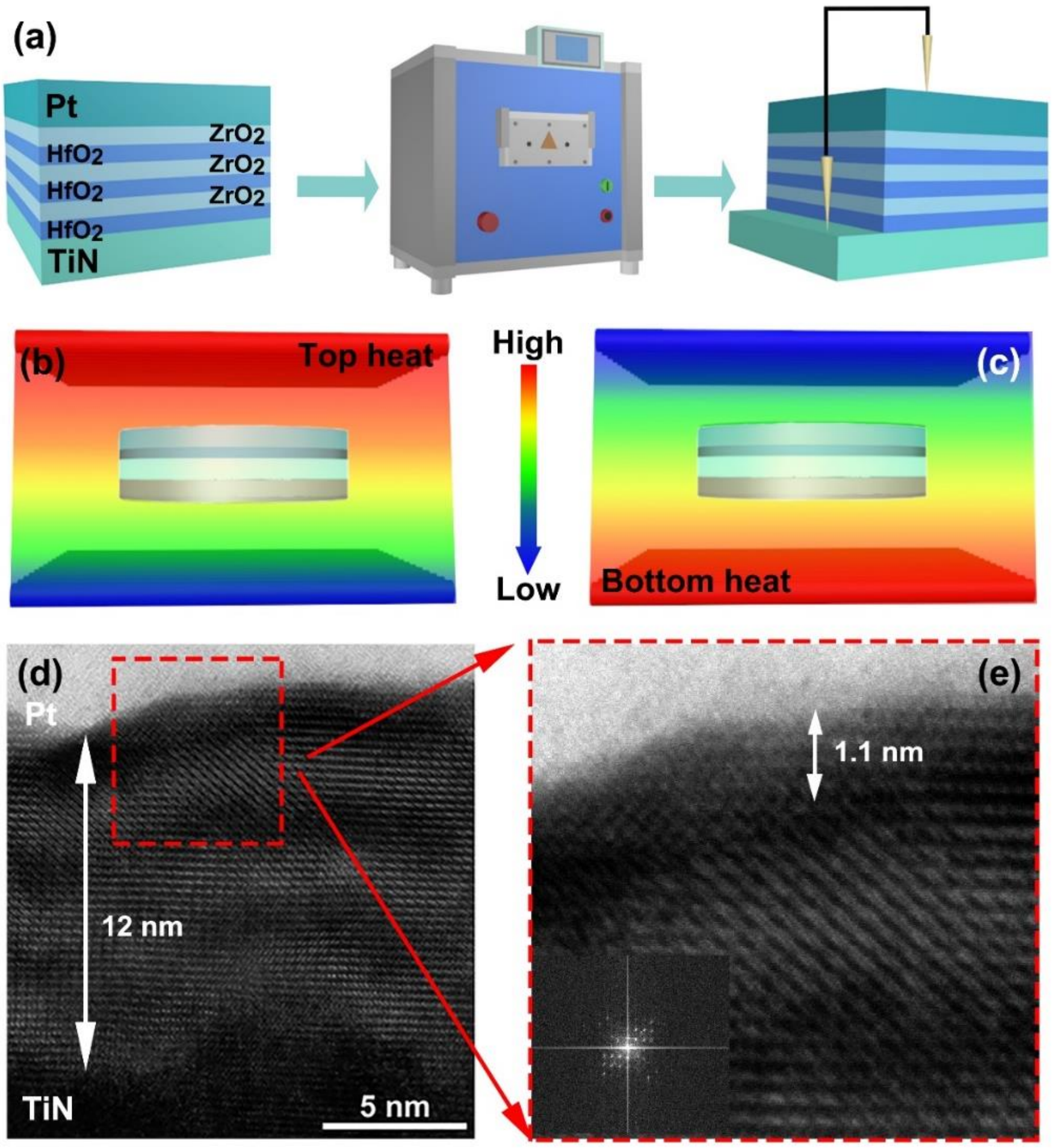

Figure 1. (a) A schematic of the capacitor fabrication—RTA—measurement process, illustrating the capacitor structure; (b) Temperature profile of the top heating mode in RTA; (c) Temperature profile of the bottom heating mode in RTA; (d) Cross-section STEM-HAADF (high-angle annular dark-field) image of the TiN/HZO/Pt ferroelectric capacitor (Sample T) after annealing; (e) A reconstructed image obtained by inverse Fourier transform from the red square area of (d).

To verify the above design principles, we first attempted the crystallization annealing with two distinct modes: (a) heating by the top lamps as in Figure 1(b) (top mode, Sample T); (b) heating by the bottom lamps as in Figure 1(c) (bottom mode, Sample B). The temperature profiles during RTA are also roughly demonstrated. The red color denotes high temperatures, and conversely, blue color represents lower temperatures. When heating from the top (bottom) side, the local temperature changes from high (low) to low (high) when scanning from TE (Pt) to BE (TiN). The device consists of $HfO_2/ZrO_2/HfO_2/ZrO_2/HfO_2/ZrO_2$, each of ~2 nm thickness, as illustrated in Figure 1(d). The bottom and top layers are $HfO_2$ and $ZrO_2$, respectively.

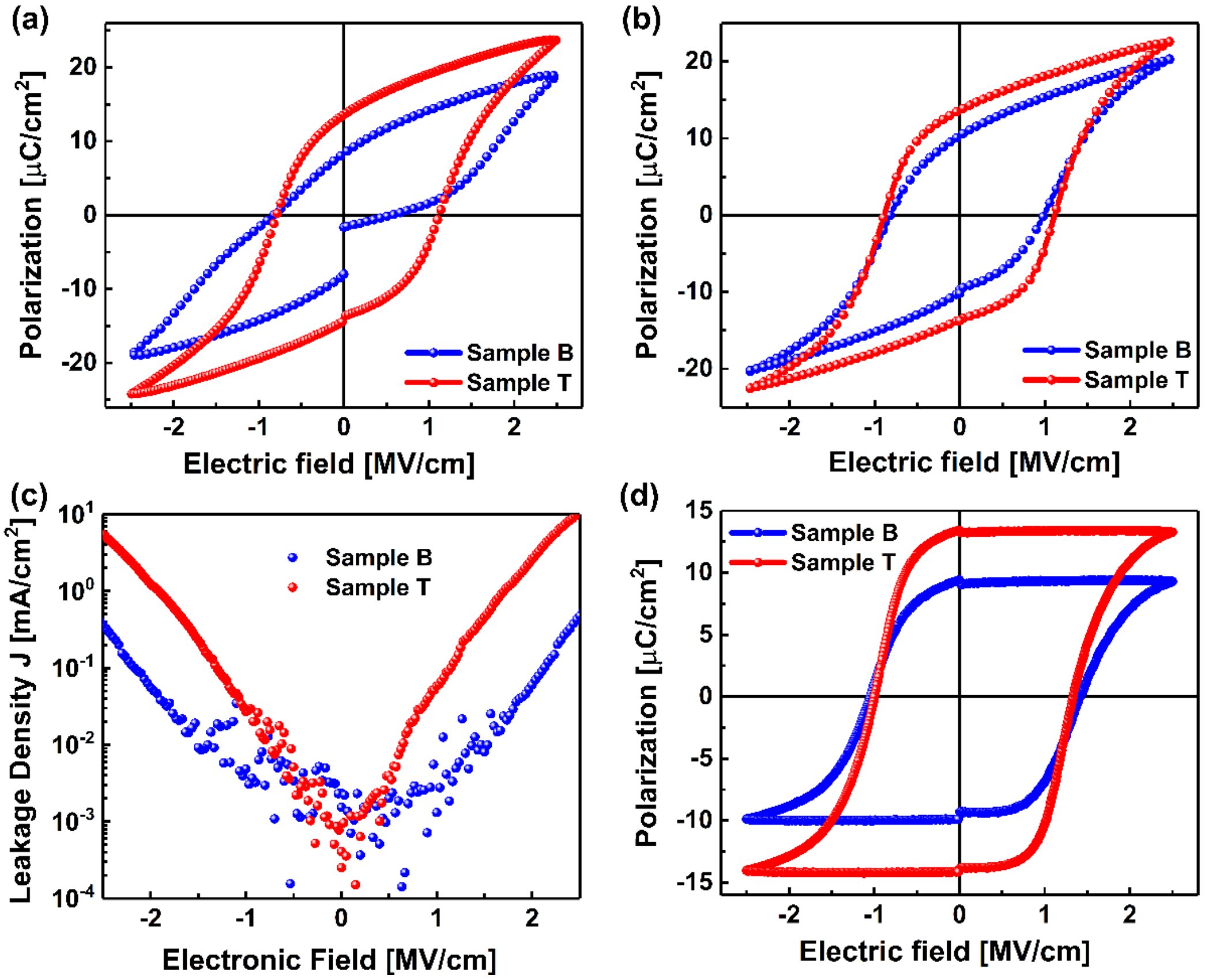

Figure 2. (a) Ferroelectric *P*-*E* hysteresis loops obtained from pristine Sample T and Sample B; (b) Ferroelectric hysteresis loops measured after the wake-up process; (c) DC leakage currents of the two samples; (d) The PUND test results.

To understand the effect of different RTA processes on ferroelectric properties, a triangular wave voltage signal with 1 kHz frequency was applied to both samples. Figure 2(a) demonstrates the pristine ferroelectric hysteresis loops, where that of Sample B is pinched. The wake-up process was subsequently applied to enhance the value of remnant polarization ($P_r$).[1, 26-28] Applying a triangular wave pulse signal 1000 times, the *P*-*E* curves become that shown in Figure 2(b). The $2P_r$ values are 27.4μC/cm$^2$ and 20.3μC/cm$^2$ for Sample T and Sample B, respectively, at 3V operation voltage. Wake-up-free characteristics have been revealed in Sample T only. The positive coercive field ($E_c^+$) of Sample T (~1.11 MV/cm) is also slightly larger than its $E_c^-$ (-0.88 MV/cm) as well as the $E_c^+$ of Sample B (~0.98 MV/cm), which is consistent with its enhanced spontaneous polarization ($P_S$), and reflects certain internal bias field in Sample T. DC leakage test results show that the strong polarization in Sample T is achieved at a cost of stronger leakage current density (9.3 mA/cm$^2$ at 2.5 MV/cm), but the overall leakage deterioration of Sample T is not severe. Furthermore, to eliminate any non-intrinsic contribution to the polarization, the positive-up-negative-down (PUND) measurement was conducted, with a double triangular wave of 1 kHz (methodology explained in Supplementary Note S1). The double triangular wave consists of a positive switching pulse, a positive non-switching pulse, a negative switching pulse and a negative non-switching pulse. The value of intrinsic $P_r^+$ is obtained from the difference between $P$(positive switching) and $P$(positive non-switching).[29-31] The value of intrinsic $P_r^-$ is obtained by the same means.[32] The rectified values of $2P_r$ are 27.13 μC/cm$^2$ and 18.58 μC/cm$^2$ for Sample T and Sample B as illustrated in Figure 2(d), confirming the intrinsic ferroelectric property of both samples.

A non-centrosymmetric $Pca2_1$ orthorhombic phase ($o$-phase) has been widely accepted as the origin of ferroelectricity in hafnia-based dielectrics.[33, 34] The impact of RTA scheme on the crystal structure of the dielectric is therefore our immediate concern. We further analyzed the crystallization of the HZO films with grazing incidence X-ray diffraction (GIXRD) characterization. As illustrated in Figure 3(a), the mixture of $o$(111) and $t$(001) peaks are located around 30.5°. The relative area ratios of the $o$-phase, $t$-phase and $m$-phase at 29°~33° are further illustrated in Figure 3(b)-3(c). The most obvious feature is that the areal intensity ratio, $o$-phase to $t$-phase, greatly increases from 1.19 as in Sample B to 6.00 as in Sample T (Table I). Moreover, the percentage of $m$-phase is enhanced in Sample T compared with Sample B. An additional $m(\bar{1}11)$ peak at 28.5° only emerges in Sample T. Hence, the major discrepancy of Sample T and Sample B lies in that the former prefers the $m$-phase together with the $o$-phase while the latter contains a substantial amount of $t$-phase.

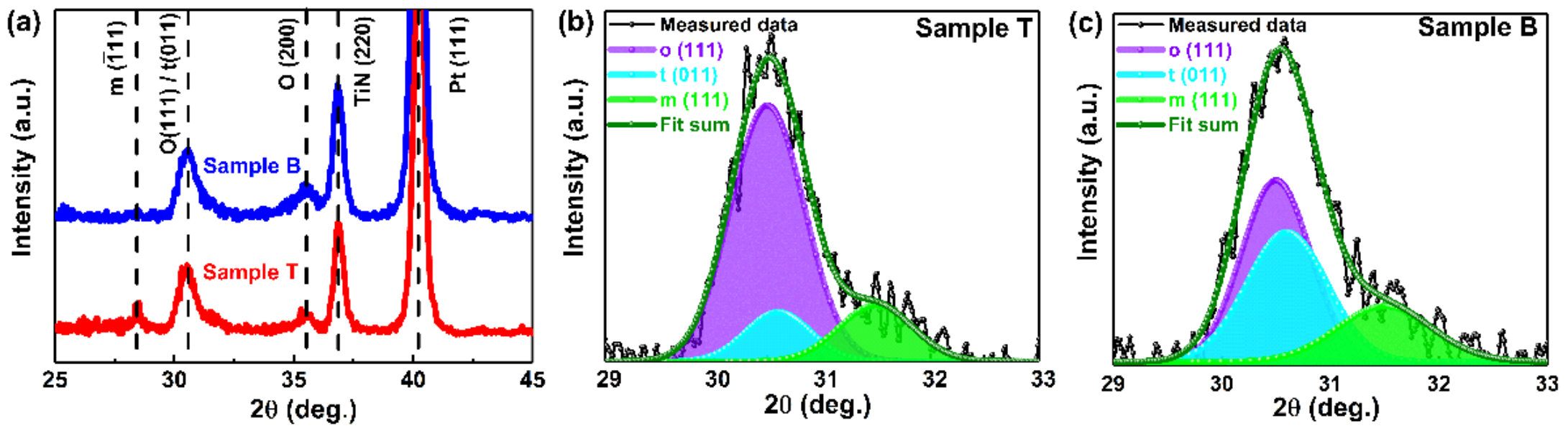

Figure 3. GIXRD characterization of the annealed ferroelectric capacitor samples. (a) GIXRD scanning results from 25° to 45° for Sample T and Sample B; (b) Phase-decomposed patterns of Sample T; (c) Phase-decomposed patterns of Sample B.

Table I. Relative areal of *o*-phase, *t*-phase and *m*-phase in Sample T and Sample B.

| Device | $A_{(o\text{-phase})}$ | $A_{(t\text{-phase})}$ | $A_{(m\text{-phase})}$ | $A_{(o\text{-phase})}/A_{(t\text{-phase})}$ |
|---|---|---|---|---|
| Sample T | 192 | 32 | 42 | 6 |
| Sample B | 135 | 113 | 57 | 1.19 |

The abundant existence of *t*-phase may explain the need of wake-up in Sample B and the possible antiferroelectric-like hysteresis loop at pristine. To further evaluate this point, we subsequently explored the distribution of the internal field of the samples. The first-order reversal curves (FORCs) were measured to reveal the contribution of the internal field by switching density.[35, 36] As shown in Figures 4(a)-4(b), the switching densities for Sample T at the first and the 1000$^{th}$ cycles are very similar. Nevertheless, two opposite internal biases appear in the pristine Sample B, located near -0.16 MV/cm and 0.79 MV/cm, respectively. The existence of double biases is relevant to the pinched hysteresis loop and antiferroelectric-like characteristics before waking-up. [1] As expected, after 1000 cycles there are only positive internal bias field emerging in Sample B, in a similar situation as Sample T.

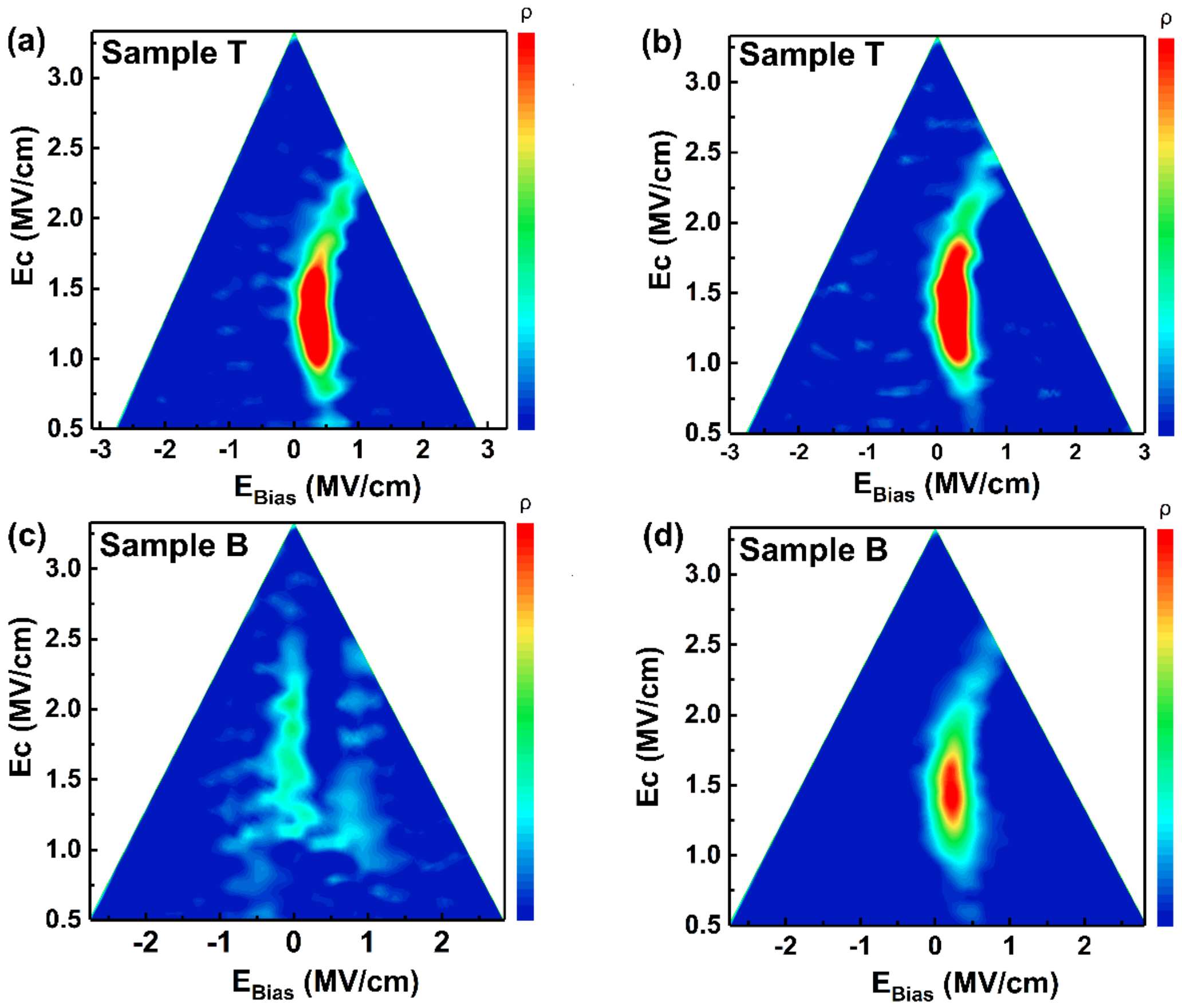

Figure 4. Distribution of the internal fields revealed by switching density in the FORCs method. (a) Sample T in its pristine state; (b) Sample T after 1000 polarization cycles; (c) Sample B in its pristine state; (d) Sample B after 1000 polarization cycles.

Based on the experimental measurements, we propose an explanation to the distinct properties between Sample T and Sample B. The location of heat source determines which side of the dielectric is first subject to crystallization. Moreover, the generation and movement of defects probably occur near the electrode that is close to the heat source. The starting point of the crystallization is important in that it can promote the emergence of similar phases in the rest part of the dielectric. It is supposed that the $ZrO_2$ layer adjacent to the TE is populated with oxygen vacancies, due to the bombardment effect of sputtering. However, the Pt TE itself is chemically inert, therefore these oxygen vacancies have a physical rather than chemical origin. When heated from the top, the generation of new oxygen vacancies is unlikely, but the existing oxygen vacancies tend to diffuse toward the BE side. A vertical distribution of oxygen vacancies facilitates the creation of conductive paths, thus increasing the leakage current. However, a potential benefit lies in that the decrease of oxygen vacancy concentration near the TE, where crystallization first occurs, can suppress the formation of the *t*-phase.[37] It is well-known that the *t*-phase is the one among $HfO_2$/$ZrO_2$ polymorphs that can tolerate the most amount of oxygen vacancies.[24,38,39] Consequently, Sample T contains less *t*-phase, but more *m*-phase. On the contrary, the TiN BE may absorb oxygen atoms at high temperature, and in Sample B the crystallization near the BE is accompanied by the generation of more oxygen vacancies locally. The formation of *t*-phase is therefore reasonable, which, together with the high oxygen vacancy concentrations near both electrode interfaces, accounts for the initially pinched hysteresis loop and the existence of waking-up. On the other hand, the leakage current of Sample B is lower as there is less driving force to create conductive paths across the entire dielectric.

**III. Comparative Study with other capacitor designs**

To further evaluate our designing principle, i.e., high temperature annealing starting from the Pt/$ZrO_2$ side, we carried out several comparative studies. First of all, we annealed the sample at 500°C using top heating mode, leading to the so-called Sample TL (L for relatively low temperature), which differs from Sample T only in terms of the annealing temperature. The initial *P-E* loop of Sample TL is pinched as shown in Supplementary Note S2, and it shows the wake-up behavior as the hysteresis gradually opens. This phenomenon is consistent with the previous reports that high annealing temperature can avoid waking-up.[17] Subsequently, we prepared three more groups of samples as demonstrated in Figure 5. The first comparative group (named Sample C1) has a reversed sequence of $HfO_2$/$ZrO_2$ growth, now with TiN/$ZrO_2$ BE interface and Pt/$HfO_2$ TE interface. Sample C2/C3 has $HfO_2$/$ZrO_2$ in contact with both electrodes (Figure 5(c)-5(d)). The chemical stoichiometry is still kept as $Hf_{0.5}Zr_{0.5}O_2$ since a corresponding $HfO_2$/$ZrO_2$ layer at the center of the dielectric possesses a reduced thickness of 1 nm for Sample C2/C3. The compositions of these comparative samples are such chosen as to examine whether heating from a Pt/$HfO_2$ interface leads to *m*- $HfO_2$-dominated phases with potentially low spontaneous polarization. All crystallization annealing steps were performed using the top heating mode at 600°C in Samples C1-C3. As shown in Figure 5(e), only Sample C3 demonstrates a low leakage current as Sample T, but Samples C1 and C2 suffer from giant leakage currents, demonstrating memristive *I*-*V* behaviors. Figure 5(e) shows that the device resistance level can turn from a low resistance state to a high resistance state when a negative voltage is present (RESET), typical of a memristor behavior. The conductive behaviors of Samples C1 and C2 have covered up their possible ferroelectric properties. On the other hand, Sample C3 exhibits a similar ferroelectric hysteresis loop as Sample T (Figure 5(f)), though the remnant polarization values are slightly inferior to Sample T.

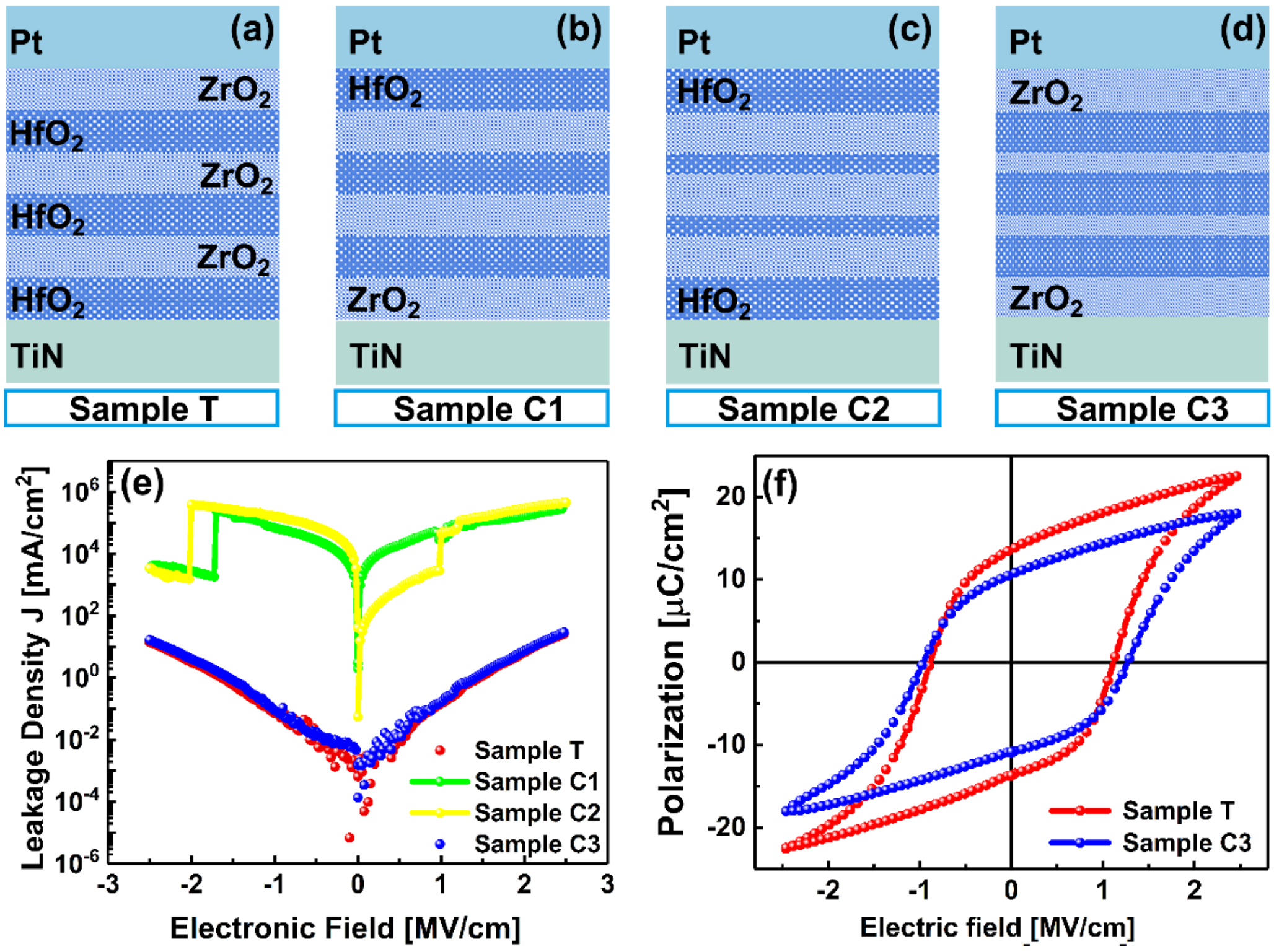

Figure 5. Comparison of the superlattice structures between (a) Sample T; (b) Sample C1; (c) Sample C2; and (d) Sample C3. (e) DC leakage current test results for all four samples; (f) Ferroelectric hysteresis loops of Sample T and Sample C3.

It follows that top heating from the Pt/$HfO_2$ interface side renders an unexpected memristive behavior, which could be understood from the different trends of oxygen vacancy clustering in $HfO_2$ compared with $ZrO_2$. We carried out first principles calculations on the energetics of filament phase segregation in $HfO_2$ and $ZrO_2$, as shown in Supplementary Note S3. The results reveal that oxygen vacancies in $HfO_2$ have a much stronger tendency to accumulate, yielding Hf or hexagonal $Hf_yO$ metal phases. The high annealing temperature not only drives the oxygen vacancies towards the TiN electrode side, but they may merge to yield metallic conductive filaments, whose growth accompanies the movement of a metallic tip towards the TiN electrode side, eventually leading to electroforming. These filaments can be subject to rupture at an applied voltage of the opposite polarity. However, for Sample T and Sample C3, formation of metallic filament is difficult in the $ZrO_2$ layer adjacent to Pt, and oxygen vacancies mainly diffuse towards the BE side, without prominent defect accumulation or phase decomposition. Therefore, these two samples do not show resistive switching, though their leakage currents are raised compared with Sample B.

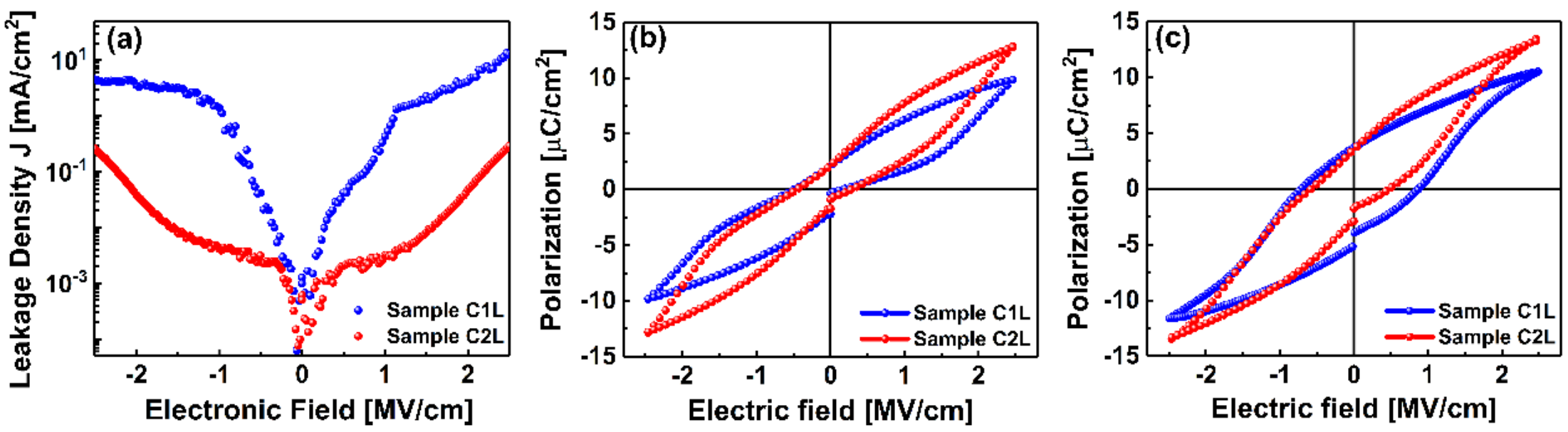

Figure 6. (a) DC leakage currents of Sample C1L and Sample C2L; (b) Ferroelectric hysteresis loops of Sample C1L and Sample C2L in their pristine states; (c) Ferroelectric hysteresis loops of Sample C1L and Sample C2L after 1000 polarization cycles.

The high 600°C annealing temperature is supposed to be a key factor for the memristor formation. Hence, we further annealed samples of the same compositions as C1 and C2, using 500°C top-heating (named Sample C1L and Sample C2L, respectively). Their leakage currents are much improved, but the initial *P*-*E* loops are strongly pinched (see Figure 6(a) and Figure 6(b)) as expected. Even after 1000 cycles, their hysteresis loops show relatively poor spontaneous polarizations, as verified by Figure 6(c).

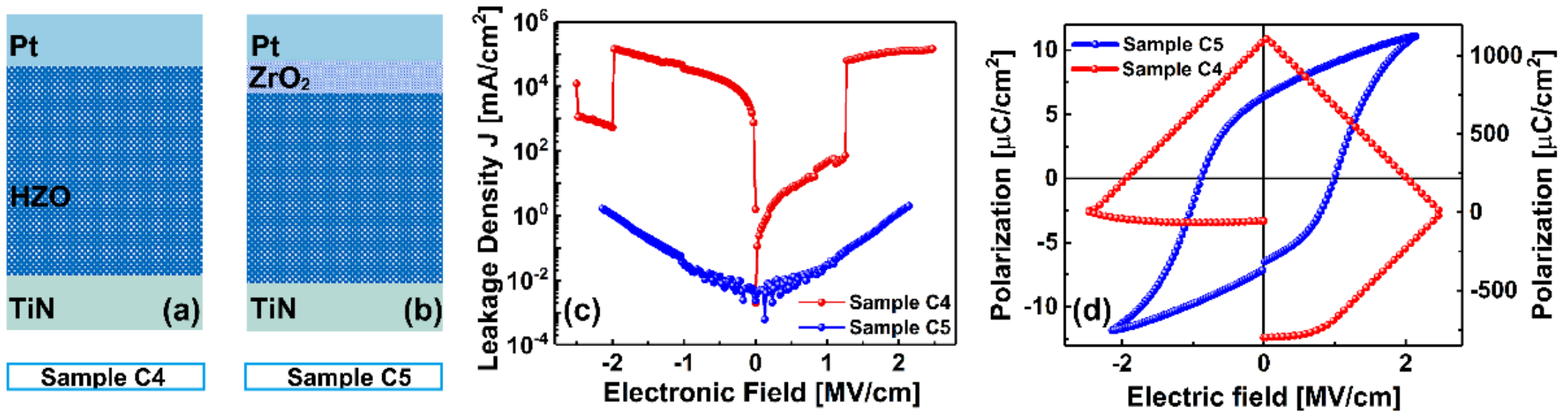

Figure 7. (a) Schematic structure of Sample C4; (b) Schematic structure of Sample C5; (c) *I-V* test results of both samples; (d) *P-E* test results of both samples.

The annealing temperature-controlled transition from a memristor to a ferroelectric capacitor requires more in-depth investigation, especially for the role of Pt/$ZrO_2$ interface in suppressing the conductive phase formation. To further strengthen or criticize these points, we fabricated ferroelectric capacitors based on $Hf_{0.5}Zr_{0.5}O_2$ solid-state solutions. That is, one cycle of $HfO_2$ deposition is followed by one cycle of $ZrO_2$, and vice versa, distinct from the superlattice growth mode. Sample C4 was prepared exactly in this way (*cf.* Figure 7(a)), and its annealing mode was still top heating at 600ºC. Figure 7(c) shows that Sample C4 already becomes a memristor with initially low resistance. However, we further fabricated an additional 2 nm-thick $ZrO_2$ layer on top of the $Hf_{0.5}Zr_{0.5}O_2$ solid-state solution, yielding the so-called Sample C5 (Figure 7(b)), whose crystallization annealing was kept the same as that of C4. Surprisingly, the leakage current of Sample C5 is much lower than that of Sample C4, by six orders of magnitude as shown in Figure 7(c). This proves the important role of Pt/$ZrO_2$ interface in avoiding the conductive filament formation at high temperature. The *P-E* loops of Samples C4 and C5 are compared in Figure 7(d). While an ill-shaped loop is discovered for Sample C4, Sample C5 exhibits a well-saturated hysteresis, indicating a typical ferroelectric behavior.

These experimental findings in general indicate that high temperature annealing can eliminate the wake-up process, provided that severe oxygen vacancy generation is avoided and the dielectric is not fully converted to the *m*-phase. We note that unidirectional heating during the RTA, *i.e.*, heating only with the top lamps, is at the heart of this technique. The unidirectional RTA is not only applicable in superlattice HZO ferroelectrics, but may be helpful in other asymmetric ferroelectric capacitor designs. It is therefore recommended to optimize the RTA process for the crystallization of other HZO-ferroelectrics in the future.

## IV. Conclusions

In conclusion, we have designed a special ferroelectric capacitor based on the $HfO_2/ZrO_2$ superlattices, which could endure high annealing temperature around 600°C and is free of waking-up. The bottom and top electrodes are TiN and Pt, respectively. What is special lies in that the dielectric layer adjacent to the Pt top electrode should be $ZrO_2$, and the rapid thermal annealing is carried out only using top lamps. The significant role of $ZrO_2$ close to the top electrode is two folds. On the one hand, it is not easily transformed to the monoclinic phase totally, even at high temperatures. On the other hand, oxygen vacancy clustering is much less probable in $ZrO_2$ than in $HfO_2$, thus avoiding the dielectric breakdown phenomenon during high temperature annealing. The Pt top electrode is preferred for its inert nature, which tends not to generate more oxygen vacancies at high temperatures. The sample annealed at 600°C shows a typical $2P_r$ value of 27.4 μC/cm$^2$, and is free of waking-up. In contrast, samples annealed at 500°C required a wake-up process. And when the Pt/$HfO_2$ contact is designed for the top electrode interface, the device becomes a memristor after 600°C annealing, which is related to the strong tendency of oxygen vacancy clustering and metal phase segregation inside $HfO_2$. The designing rule of this work as well as the high temperature annealing process for wake-up removal can be useful for the material and device design of hafnia-based ferroelectrics.

## Experimental

The TiN/$HfO_2$-$ZrO_2$ superlattice/Pt device is fabricated on an n-Si/$SiO_2$ substrate. The bottom electrode TiN was deposited by magnetron sputtering. Subsequently, the 12nm-

$HfO_2$-$ZrO_2$ superlattice film was further deposited by thermal atomic layer deposition (ALD) at 300℃. The 100nm-Pt top electrode of an area 50×50μ$m^2$ were patterned on the HZO film by lithography and lift-off process. Finally, the two mode RTA processes were carried out at 600°C in a nitrogen atmosphere for 30s. Setting the annealing mode as (i) heating from the top lamp (Sample T). (ii) heating from the bottom lamp, (Sample B), respectively.

The thickness of the device was characterized by Cross section TEM. the crystal structure of the TiN/$HfO_2$-$ZrO_2$ superlattice/Pt device were obtained by using grazing incidence X-ray diffraction (GIXRD). The P-E curves of the devices were measured with an Agilent B1530 semiconductor analyzer at room temperature.

**Supporting Information**

Supporting Information is available.

**Acknowledgements**

This work was supported by the Funding: National Key Research and Development Program of China(2021ZD0114401) and the National Natural Science Foundation of China under Grant No. 61974049 and 61974047.

**Conflict of Interest**

The authors declare no conflict of interest.

Supplementary Information for

# Designing wake-up free ferroelectric capacitors based on the $HfO_2/ZrO_2$ superlattice structure

*Na Bai,*[1] *Kan-Hao Xue,*[1,2*] *Jinhai Huang,*[1] *Jun-Hui Yuan,*[1] *Wenlin Wang,*[1] *Ge-Qi Mao,*[1] *Lanqing Zou,*[1] *Shengxin Yang,*[1] *Hong Lu,*[1] *Huajun Sun,*[1,2*] *and Xiangshui Miao*[1,2]

[1] School of Integrated Circuits, School of Optical and Electronic Information, Huazhong University of Science and Technology, Wuhan 430074, China

[2] Hubei Yangtze Memory Laboratory, Wuhan 430074, China

## Supplementary Note S1

The positive-up-negative-down (PUND) measurement was conducted, with the frequency and amplitude of the double triangular wave are 1 kHz and 3V. The double triangular wave consists of a positive switching pulse, a positive non-switching pulse, a negative switching pulse and a negative non-switching pulse. The value of intrinsic $P_r^+$ is obtained from the difference between *P*(positive switching) and *P*(positive non-switching) [1-3]

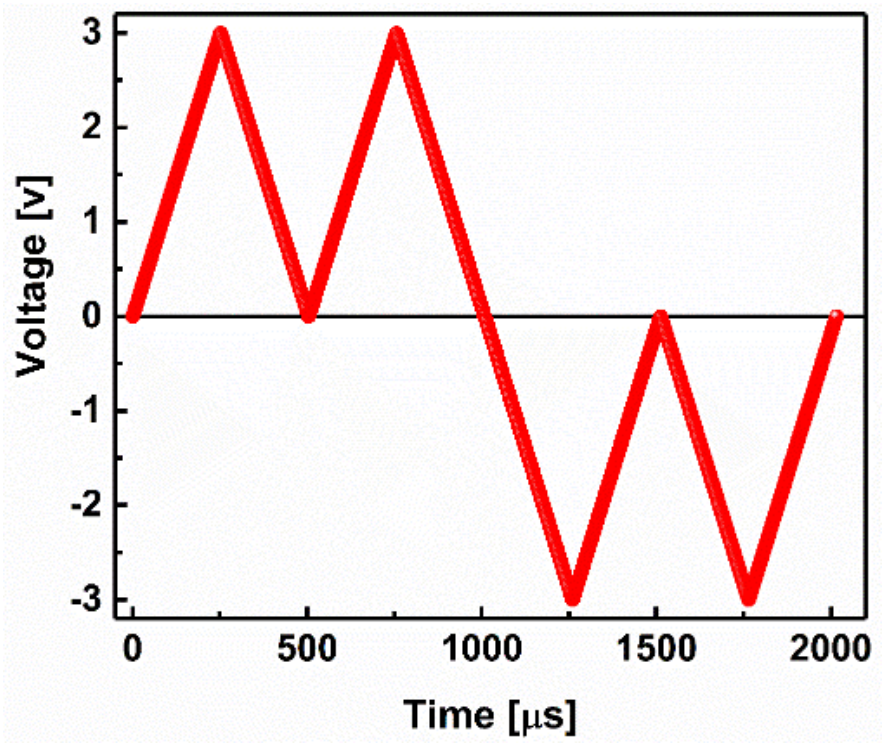

Figure S1. The double triangular wave used for the positive-up-negative-down (PUND) measurement.

## Supplementary Note S2

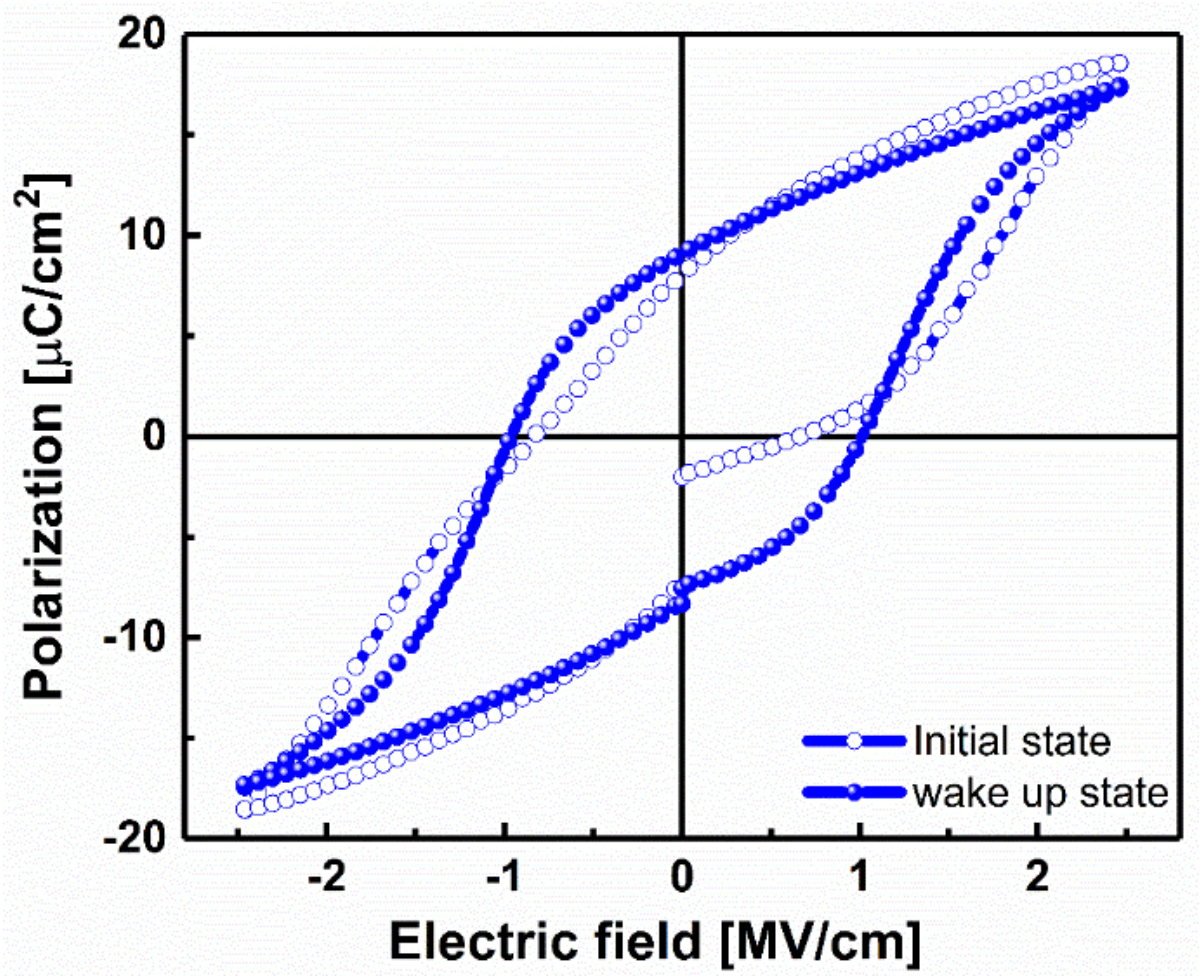

Figure S2. Ferroelectric hysteresis loops of Sample TL measured from initial state and wake up state.

We annealed the sample T at 500°C using top heating mode, named Sample CTL, which differs from Sample T only in terms of the annealing temperature. The initial *P*-*V* loop of Sample CTL is pinched as shown in Figure S2, and it shows the wake-up behavior as the hysteresis gradually opens. This phenomenon is consistent with the previous reports that high annealing temperature can avoid waking-up[4]

**Supplementary Note S3**

The highly conductive filament formation in $HfO_2$ has been shown to accompany a phase decomposition into hexagonal metal phases like metal Hf , such as $h$-$Hf_6O$. The other Hf suboxides $HfO_y$ with $y \geq 1$ are typically semiconductors or semi-metals with relatively poor conductivity. Here we use first-principles calculations to evaluate the energy cost of such phase decomposition with hexagonal metallic phases in defective $HfO_x$ and $ZrO_x$.

The density functional theory (DFT) calculations were carried out using plane-wave based Vienna *Ab initio* Simulation Package (VASP 5.4.4). A 500 eV energy cutoff was set to truncate the plane-wave basis. The generalized gradient approximation (GGA) was adopted for the exchange-correlation energy, within the Perdew-Burke-Ernzerhof (PBE) functional form. The state-of-the-art projector augmented-wave method was used. The electrons considered in the valence are: 5p, 5d and 6s for Hf, 2s and 2p for O, 4s, 4p, 4d and 5s for Zr. Table S1 lists the calculated materials and their total energies, which were obtained though full structural optimization followed by static total energy calculations.

Based upon these energy values, the energy cost of phase decomposition from a defective Hf/Zr sub-oxide to hexagonal phases can readily be estimated. Table S2 demonstrates the corresponding results. It is observed that Hf sub-oxides are much more easily decomposed to yield hexagonal phases for possible filament formation, compared with Zr sub-oxides. For instance, the decomposition of $Hf_2O_3$ into Hf has a energy gain of 0.53 eV/f.u., but that of $Zr_2O_3$-to-Zr decomposition only exhibits a tiny energy gain of 0.01 eV/f.u. Moreover, hexagonal ZrO is demonstrated to be stable against decomposition into Zr or $Zr_6O$, which is however not the case for the lowest energy phase of HfO. These results reveal that in $HfO_x$ there is a strong trend of oxygen vacancy cohesion to yield highly conductive filaments, compared with $ZrO_x$.

**Table S1**. Materials under investigation, the corresponding calculation settings, and results

| Material | Computational | | | | |
|---|---|---|---|---|---|
| | Number of atoms per cell | | | K-mesh in sampling the Brillouin zone | Total energy per cell (eV) |
| | Hf | Zr | O | | |
| Hf | 2 | 0 | 0 | 15×15×11 | -19.85 |
| HfO ($I4_1/amd$) | 16 | 0 | 16 | 9×9×5 | -321.80 |
| $m$-$HfO_2$ unit cell | 4 | 0 | 8 | 9×9×9 | -122.08 |
| $m$-$HfO_2$ supercell | 24 | 0 | 48 | 5×3×10 | -732.47 |
| $Hf_2O_3$ ($P\bar{4}m2$) | 32 | 0 | 48 | 5×5×6 | -803.42 |
| $h$-$Hf_6O$ | 18 | 0 | 3 | 15×15×11 | -210.55 |
| Zr | 0 | 2 | 0 | 9×9×5 | -17.04 |
| ZrO ($P\bar{6}2m$) | 0 | 16 | 16 | 9×9×9 | -298.99 |
| $m$-$ZrO_2$ unit cell | 0 | 4 | 8 | 5×3×10 | -115.04 |
| $m$-$ZrO_2$ supercell | 0 | 24 | 48 | 5×6×6 | -690.23 |
| $Zr_2O_3$ ($P\bar{4}m2$) | 0 | 32 | 48 | 9×9×3 | -758.23 |
| $h$-$Zr_6O$ | 0 | 18 | 3 | 15×15×11 | -184.84 |

**Table S2**. Energetics of various sub-oxide decomposition reactions. Here f.u. represents a formula unit.

| Reaction | $E_1$ (eV/f.u.) | $E_2$ (eV/f.u.) | $E_3$ (eV/f.u.) | $\Delta E = (E_2+E_3)-E_1$ (eV/f.u.) |
|---|---|---|---|---|
| $HfO \rightarrow \frac{1}{2} Hf + \frac{1}{2} HfO_2$ | $HfO$ | $\frac{1}{2} Hf$ | $\frac{1}{2} HfO_2$ | -0.11 |
| | -20.11 | -4.96 | -15.26 | |
| $HfO \rightarrow \frac{5}{11} HfO_2 + \frac{1}{11} Hf_6O$ | $HfO$ | $\frac{5}{11} HfO_2$ | $\frac{1}{11} Hf_6O$ | -0.14 |
| | -20.11 | -13.87 | -6.38 | |
| $Hf_2O_3 \rightarrow \frac{1}{2} Hf + \frac{3}{2} HfO_2$ | $Hf_2O_3$ | $\frac{1}{2} Hf$ | $\frac{3}{2} HfO_2$ | -0.53 |
| | -50.21 | -4.96 | -45.78 | |
| $Hf_2O_3 \rightarrow \frac{16}{11} HfO_2 + \frac{1}{11} Hf_6O$ | $Hf_2O_3$ | $\frac{16}{11} HfO_2$ | $\frac{1}{11} Hf_6O$ | -0.56 |
| | -50.21 | -44.39 | -6.38 | |
| $ZrO \rightarrow \frac{1}{2} Zr + \frac{1}{2} ZrO_2$ | $ZrO$ | $\frac{1}{2} Zr$ | $\frac{1}{2} ZrO_2$ | 0.05 |
| | -18.69 | -4.26 | -14.38 | |
| $ZrO \rightarrow \frac{5}{11} ZrO_2 + \frac{1}{11} Zr_6O$ | $ZrO$ | $\frac{5}{11} ZrO_2$ | $\frac{1}{11} Zr_6O$ | 0.02 |
| | -18.69 | -13.07 | -5.60 | |
| $Zr_2O_3 \rightarrow \frac{1}{2} Zr + \frac{3}{2} ZrO_2$ | $Zr_2O_3$ | $\frac{1}{2} Zr$ | $\frac{3}{2} ZrO_2$ | -0.01 |
| | -47.39 | -4.26 | -43.14 | |
| $Zr_2O_3 \rightarrow \frac{16}{11} ZrO_2 + \frac{1}{11} Zr_6O$ | $Zr_2O_3$ | $\frac{16}{11} ZrO_2$ | $\frac{1}{11} Zr_6O$ | -0.04 |
| | -47.39 | -41.83 | -5.60 | |